\documentclass[aps,prl,twocolumn,superscriptaddress,groupedaddress]{revtex4}  

\usepackage{times}
\usepackage{graphicx}  
\usepackage{dcolumn}   
\usepackage{mathtools} 
\usepackage{dsfont}
\usepackage{bm}
\usepackage{bbm}     
\usepackage{amssymb}   
\usepackage{siunitx}
\usepackage[breaklinks=true,colorlinks,citecolor=blue,linkcolor=blue,urlcolor=blue]{hyperref}
\usepackage{multirow}
\usepackage{xcolor}
\usepackage{amsmath}
\usepackage{hhline}
\usepackage{lipsum}

\usepackage{color} 

\begin{document}

\title{Feature Spectrum Topology} 

\author{Baokai Wang}
\thanks{B. W. and Y. H. contributed equally to this work.}
\affiliation{Department of Physics, Northeastern University, Boston, Massachusetts 02115, USA}

\author{Yi-Chun Hung}
\thanks{B. W. and Y. H. contributed equally to this work.}
\affiliation{Department of Physics, Northeastern University, Boston, Massachusetts 02115, USA}
\affiliation{Institute of Physics, Academia Sinica, Taipei 115201, Taiwan}
\affiliation{Department of Physics, National Taiwan University, Taipei 106216, Taiwan}

\author{Xiaoting Zhou}
\affiliation{Department of Physics, Northeastern University, Boston, Massachusetts 02115, USA}

\author{Tzen Ong}
\affiliation{Institute of Physics, Academia Sinica, Taipei 115201, Taiwan}

\author{Hsin Lin}
\email{nilnish@gmail.com}
\affiliation{Institute of Physics, Academia Sinica, Taipei 115201, Taiwan}


\begin{abstract}
Topology is a fundamental aspect of quantum physics, and it has led to key breakthroughs and results in various fields of quantum materials. In condensed matters, this has culminated in the recent discovery of symmetry-protected topological phases. However, symmetry-based topological characterizations rely heavily on symmetry analysis and are incapable of detecting the topological phases in systems where the symmetry is broken, thus missing a large portion of interesting topological physics. Here, we propose a new approach to understanding the topological nature of quantum materials, which we call feature spectrum topology. In this framework, the ground-state is separated into different partitions by the eigenspectrum of a feature, a particular chosen internal quantum degree of freedom, such as spin or pseudo-spin, and the topological properties are determined by analysis of these ground-state partitions. We show that bulk-boundary correspondence guarantees gapless spectral flows in either one of the energy or feature spectrum. Most importantly, such `feature-energy duality' of gapless spectral flows serves as a fundamental manifestation of a topological phase, thereby paving a new way towards topological characterizations beyond symmetry considerations. Our development reveals the topological nature of a quantum ground state hidden outside symmetry-based characterizations, hence, providing a platform for a more refined search of unconventional topological materials.
\end{abstract}

\maketitle

\section*{I. Introduction}
The study of ground state topology of quantum systems has been an important field in condensed matter physics, starting with the integer quantum Hall effect (IQHE) and culminating in the discovery of the quantum spin Hall effect (QSHE) and $\mathds{Z}_2$ topological insulator (TI), and the corresponding classification of an entire family of symmetry-protected topological (SPT) materials \cite{RevModPhys.88.021004, RevModPhys.82.3045, RevModPhys.83.1057}. Currently, SPT systems and materials are efficiently categorized and predicted using symmetry-based frameworks, such as topological quantum chemistry and symmetry indicators \cite{Bradlyn2017-pj, Po2017, PhysRevX.7.041069, Song2018-vx, Tang2019-wx, Vergniory2019-xq, Zhang2019-ze}, which accurately captures the band inversions at high-symmetry points and the type of topological invariant. However, this framework is inapplicable in the presence of symmetry-breaking perturbations or generic band inversions at non-symmetry-related $k$-points. Hence, we are led to consider the possibility of characterizing the non-trivial topological nature of quantum systems in the absence of protecting symmetries. 

The close relationship between the QSHE and 2D $\mathds{Z}_2$ TI provided a key insight during the formulation of this non-symmetry protected (nSPT) topological characterization \cite{PhysRevLett.97.036808, PhysRevB.75.121403, PhysRevB.80.125327, PhysRevLett.107.066602, Prodan_2011}, especially the recent advancement in characterizing spin-Chern insulators with broken spin-$U(1)$ symmetry or time-reversal symmetry using projective operators \cite{PhysRevB.80.125327,https://doi.org/10.48550/arxiv.2207.10099,https://doi.org/10.48550/arxiv.1011.5456, PhysRevLett.107.066602,PhysRevLett.108.196806, YANG2018723}. In general, the standard paradigm for understanding topology is viewed through the lens of the band structures in the energy domain. The topological invariants characterizing these topological states can be understood as arising from the Berry phase of the occupied wave function, i.e., the geometric evolution of the internal quantum degrees of freedom of the electron. In this picture, the topology is embedded in the fiber bundle constituted by the wavefunctions of valence electrons, i.e., $P\hat{H}P$. Here, $P$ is the projection operator to the occupied space, and $\hat{H}$ is the system's Hamiltonian. However, electrons in solids are characterized by all the internal quantum degrees of freedom, including spin, pseudo-spin, and lattice symmetry-induced characteristics (e.g., Mirror operator $\hat{M}_z$). This leads us to consider using the projection of a particular quantum number ($\hat{O}$) to partition the occupied electronic states, thereby obtaining the spectrum of the operator, $P \hat{O} P$. The intrinsic topology of each sector originates from the non-trivial Berry curvature of the underlying quantum number, and the overall topological phase of the entire system is characterized by all the constituent sectors, suggesting that the set of topological characters of the different sectors is a more fundamental building block underlying the topology of a quantum system. In sharp contrast to symmetry-protected topological phases, these feature topological invariants are robust in the presence of perturbative symmetry-breaking fields while providing richer topological information. We term this characterization scheme `\emph{feature spectrum topology}' and introduce it in detail below. The conventional SPT phases can be well incorporated into the feature spectrum topology and can be assigned a topological invariant from the perspective of feature spectrum topology (see Fig.~\ref{fig1}(\textbf{a})).

\section*{II. Feature Spectrum Topology}
\begin{center}
\begin{figure}[htbp]
\includegraphics[scale=0.35]{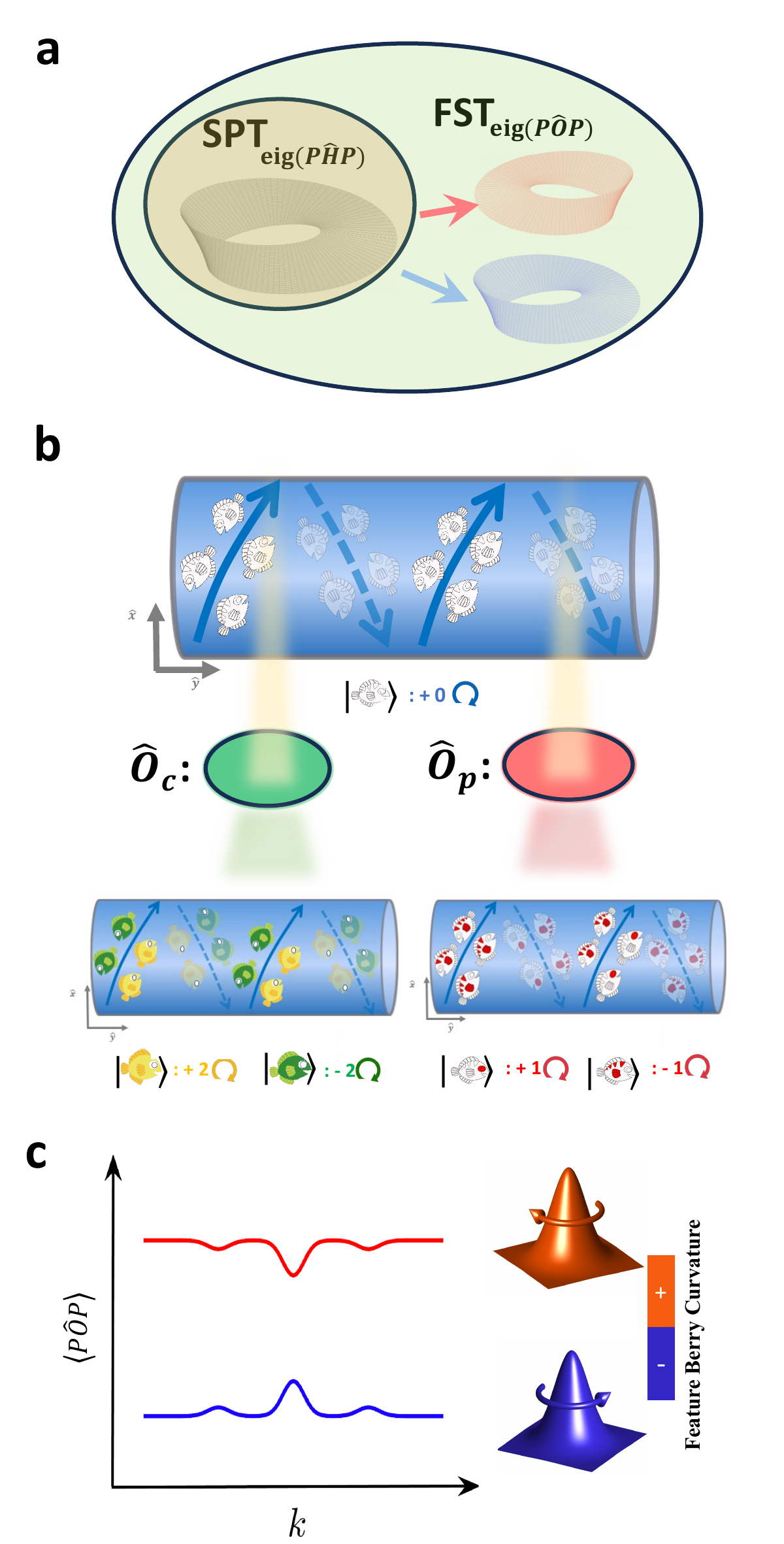}
\caption{\label{fig1}\textbf{Feature Spectrum Topology.}
(\textbf{a}) Symmetry-protected topology (SPT) focuses on the occupied states in the energy domain, i.e., $\text{eig}(P\hat{H}P)$. In contrast, feature spectrum topology (FST) separates the ground state into different sectors via the eigenspectrum of the projected quantum number, i.e. $\text{eig}(P\hat{O}P)$, and analyzes the sectors to classify the ground state topology.
(\textbf{b}) The quantum states in the Brillouin zone, like fishes living in a water tank, demonstrate their non-trivial topology through their winding number. The feature spectrum acts like a filter, reflecting the topological nature that is only manifested through the behavior of the fish subgroup, which is invisible to the whole fish group. As shown in the diagram, the whole group of fish does not have a net winding number. However, each subgroup of fish, partitioned by their \emph{features} like body color $\hat{O}_c$ or patterns $\hat{O}_p$, can carry a non-vanishing winding number. If the fishes are partitioned by their body color, the yellow fishes have a net winding number of $2$ while the green fishes have a net winding number of $-2$. If the fishes are partitioned by their pattern, the fishes with red eyes have a net winding number of $1$ while the fishes with red back strips have a net winding number of $-1$.
(\textbf{c}) The feature $\hat{O}$, representing a chosen internal quantum degree of freedom such as spin $\hat{S}_z$, separates the occupied space into sectors. The separation is demonstrated by the \emph{feature spectrum} $\langle P\hat{O}P \rangle$, on which the bands in each sector can carry Berry curvatures manifesting non-trivial ground state topology.}
\end{figure}
\end{center}

An electron in solids carries multiple properties, such as spin, orbital angular momentum, and sub-lattice index (pseudo-spin), that can be used to categorize electronic states. Formally, we denote their corresponding quantum operator $\hat{O}$ a \emph{`feature'} and $P\hat{O}P$ a \emph{`feature operator'} for convenience. Note that the choice of the feature is not unique. A schematic diagram for the feature spectrum topology is illustrated in Fig.~\ref{fig1}(\textbf{b}). Diagonalization of the projected operator produces the feature spectrum consisting of `\emph{sectors}', indexed by $n$, with feature spectral values $O_n(\mathbf{k})$ at every $k$-point, and its corresponding eigenfunction $\tilde{\psi}_{O_n}$. It should be noted that the feature is ill-defined when the conduction band and the valence band meet at degenerate points, where the occupied states cannot be uniquely defined.

The feature operator $P\hat{O}P$ partitions the occupied states into sectors according to its eigenspectrum. When $\hat{O}$ is a symmetry of the system, $\{\tilde{\psi}_{O_n} \vert \forall n\}$ are eigenstates of the Hamiltonian. In that case, the feature $P\hat{O}P$ sorts the occupied states into sectors according to their eigenvalues of the symmetry $\hat{O}$, and the feature spectrum consists of flat bands corresponding to the eigenvalues of the symmetry $\hat{O}$. In more general situations when $\hat{O}$ is not a symmetry, $\{\tilde{\psi}_{O_n} \vert \forall n\}$ are linear combinations of the eigenstates of the Hamiltonian instead. The feature spectrum consists of dispersive bands with their eigenvalues deviating from the eigenvalues of the symmetry $\hat{O}$ (see Fig.~\ref{fig1}(\textbf{c})). However, the topology of each sector can still be characterized by calculating the Wilson loop for each feature sector as long as different sectors remain separated\cite{Zuo_2015, Deng2016},
\\
\begin{equation}
W[\gamma]_O = Tr \left[ \mathcal{P}  \, exp \left( i \oint_{\gamma} \tilde{A}_{\mu} dx^{\mu}  \right)   \right].
\end{equation}
\\
Here, the Berry connection for the feature spectrum is given by $\tilde{A}_{\mu} = i \langle \tilde{\psi}_{O_n} \vert \nabla_{\mu} \vert \tilde{\psi}_{O_n} \rangle$. The feature spectrum Wilson loop, $W[\gamma]_O$, now allows us to study the topology of the spectral flow of different quantum degrees of freedom in the ground state. 

\section*{III. The Feature-Energy Duality}
A non-zero winding number, $\mathds{Z}_O$, for $W[\gamma]_O$ identifies a non-trivial topology corresponding to the feature $\hat{O}$ in the quantum ground state. Since the lab-reference frame is a topologically-trivial vacuum, bulk-boundary correspondence indicates that there have to be gapless modes in either the energy spectrum or the feature spectrum at the boundary to connect the two topologically distinct spaces. We term this result `\emph{feature-energy duality}', which is the most important outcome of this work. We discuss consequences of this feature-energy duality and how it is reflected in the bulk-boundary correspondence of feature-topological systems and also during a topological quantum phase transition in the following section.

The feature-energy duality is concisely captured in the following scenario: a SPT phase hosts gapless energy bands on the boundary when the protecting symmetry is intact. When a weak symmetry-breaking perturbation is applied, the gapless energy bands become gapped, but the topological invariants in the feature spectrum topology remain unchanged, and in this case, the feature spectrum on the boundary becomes gapless. That is, either the energy or feature spectrum on the edge will have to be gapless for a system to be in a topologically non-trivial phase. Clearly, SPT phases can be described by topological invariants of the feature spectrum, and can thus be included within this new feature spectrum topology framework.

\begin{center}
\begin{figure}[htbp]
\includegraphics[scale=0.38]{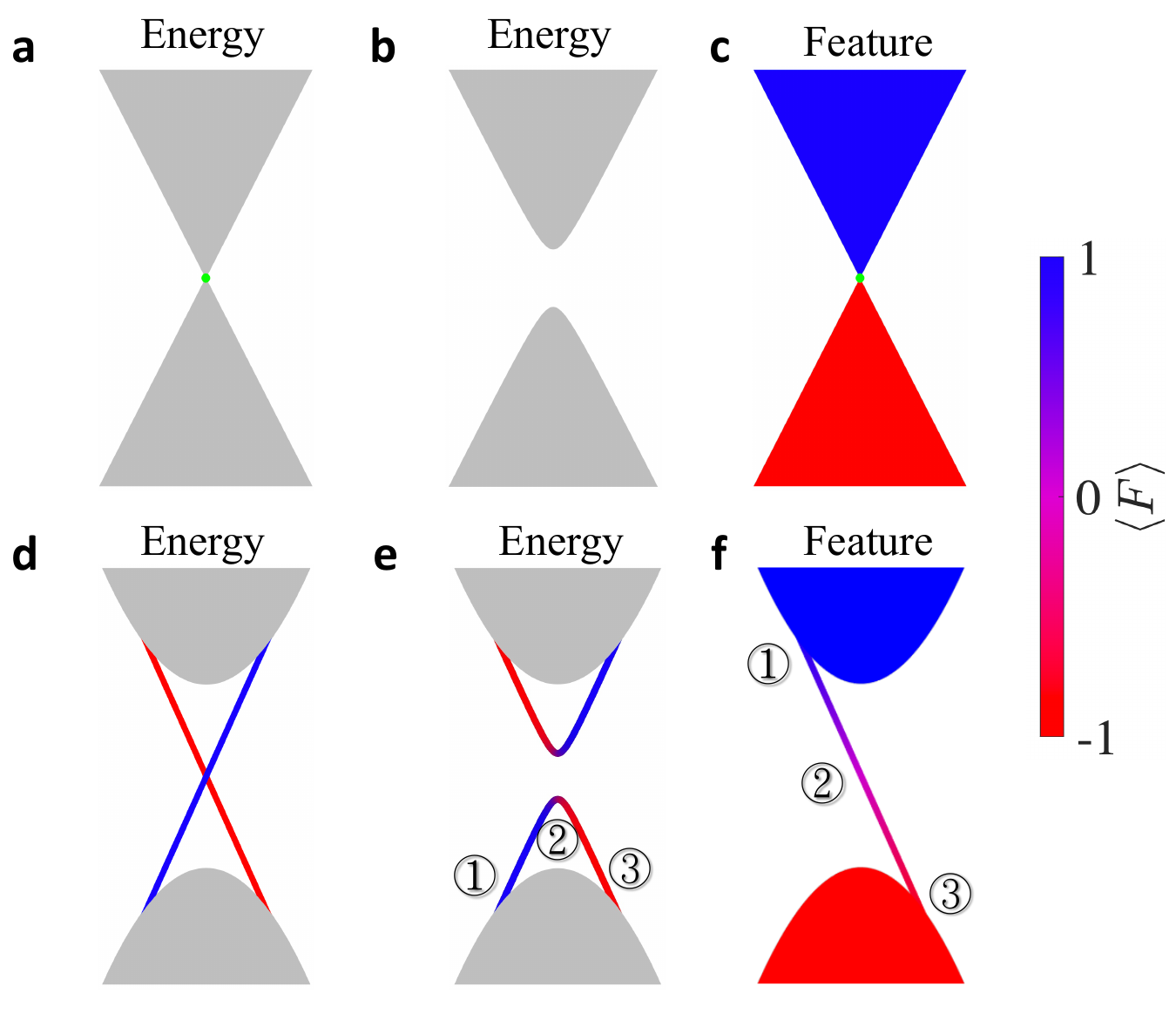}
\caption{\label{fig2} \textbf{Feature-Energy Duality.} A topological phase transition can occur in two ways. One way is to close the bulk energy band gap as shown in (\textbf{a}). The other way is to maintain the bulk energy band gap (\textbf{b}) while closing the bulk feature band gap (\textbf{c}). Feature-energy duality dictates that the bulk-boundary correspondence manifestation of the non-trivial bulk topology can occur in two ways. One way is through the appearance of gapless boundary states in the energy spectrum, connecting the valence bands and conduction bands, as shown in (\textbf{d}). The other way is to have gapped boundary states in the energy spectrum(\textbf{e}), but exhibit gapless edge states in the feature spectrum as shown in (\textbf{f}). Furthermore, the feature spectral flow reflects how the boundary states connect different feature sectors in the feature spectrum. The evolution marked by $\textcircled{1} \sim \textcircled{3}$ shows how the texture of the feature, such as spin texture, changes on the boundary energy and feature spectrum. As shown in the diagram, the feature texture gradually changes from the $+1$ sector (\textcircled{1}; blue) to the intermediate state with the feature spectral value of $0$ (\textcircled{2}; magenta), and finally to the $-1$ sector (\textcircled{3}; red). }
\end{figure}
\end{center}

\subsection*{A. Topological Quantum Phase Transition}
The feature spectrum topology extends the concept of topology from the bulk energy band structure to the bulk feature spectrum. Correspondingly, a topological phase transition may occur via the emergence of gapless modes in the bulk feature spectrum rather than just in the bulk band structure. More specifically, a distinctive signature of a topological phase is the inability to adiabatically deform its ground state to the trivial phase. 
Distinctly different from the typical topological quantum phase transition (TQPT) (Fig.~\ref{fig2}(\textbf{a})) induced by closing of the bulk energy band gap, feature spectrum topology allows for a topological phase transition to occur via gap closure in the bulk feature spectrum while keeping the bulk energy bands gapped (Fig.~\ref{fig2}(\textbf{b, c})). Similar to how an energy gap in the band structure protects the topology of SPT states, the gap in the bulk feature spectrum prevents the topology of each feature sector from deforming adiabatically to the trivial case.
Therefore, feature spectrum topology allows for a TQPT to occur via two distinctly different routes, the closing of a gap in either the energy band structure or the bulk feature spectrum (Fig.~\ref{fig2}(\textbf{c})); with the corresponding emergence of gapless nodes in the bulk feature spectrum instead of gapless nodes in the bulk energy spectrum\cite{Yang2013_edge_states}.(See Supplementary Discussion and Fig.~S2.)  

\subsection*{B. Bulk-Boundary Correspondence}
In feature spectrum topology, topological phase transition can occur via the gap closing in the energy or feature band structures. Correspondingly, the bulk-boundary correspondence manifests either through the gapless energy band structures or gapless feature spectrum on the boundaries (Fig.~\ref{fig2}(\textbf{d-f})). The occupied energy states are partitioned in the feature spectrum into different feature sectors, which can be regarded as distinct topological objects sharing the same Brillouin zone. These feature sectors may carry nontrivial topological properties; hence, bulk-boundary correspondence will also apply to the feature spectrum.
The non-trivial topology carried by a sector manifests through the existence of a gapless feature spectrum on the boundary when the boundary energy band structure is gapped (Fig.~\ref{fig2}(\textbf{e, f})), indicating the existence of boundary states connecting different feature sectors.
The expectation value of the feature in the boundary energy band gradually changes from one sector to another as the feature texture evolves in the gapless boundary feature bands (Fig.~\ref{fig2}(\textbf{e, f})).  

The standard SPT classification is inapplicable in the presence of a symmetry-breaking field, such as applying an out-of-plane electric field to a mirror Chern insulator (See Supplementary Discussion and Fig.S4). However, the feature spectrum allows us to study and distinguish the non-trivial ground state topology in such quantum systems from the trivial lab vacuum, in which gapless modes emerge in the boundary feature spectrum to connect the two topologically-inequivalent ground states. When the symmetry-breaking field is switched off, the standard SPT classification applies again, and the corresponding gapless modes are restored in the boundary energy band structure (Fig.~\ref{fig2}(\textbf{d})). Note that the feature spectrum is ill-defined at the $k$-point corresponding to the Dirac node in the boundary band structure, thereby allowing a transition between the non-trivial bulk state and the trivial vacuum. We term this simultaneous opening/ closing of a gap in the energy/ feature spectrum on the boundary a `\emph{feature-energy duality}'. The feature-energy duality exceeds the previous generalized Laughlin argument in the quantum spin Hall effect \cite{PhysRevLett.108.196806, YANG2018723} by providing more insight into the origin of the gapless boundary spectral flows in the feature spectrum, enabling applications to broader cases of unconventional topological materials.

\section*{IV. Feature Chern Insulators}
\paragraph{} When the ground states are partitioned by the feature spectrum, each feature sector is regarded as distinct topological objects. If the feature sectors can be characterized by a Chern number, we term this system a \emph{`feature Chern insulator'}\cite{PhysRevB.78.195125}. We call these integers \emph{the feature Chern numbers} in an analogy to the spin Chern numbers in which $\vec{S}\cdot\hat{n}$ is used as the feature \cite{PhysRevB.80.125327,https://doi.org/10.48550/arxiv.2207.10099,https://doi.org/10.48550/arxiv.1011.5456,PhysRevLett.108.196806, YANG2018723}. Further, the feature spectrum topology exhibits more details about the topological nature of the system, which are invisible to the traditional topological characterization method. Here, we take a high-pseudo-spin Chern insulator to demonstrate the feature Chern insulator.

\subsection*{(High-)Pseudo-Spin-Chern Insulators}
A high-pseudo-spin Chern insulator can be constructed by staking of 2D $\mathds{Z}_2$ TI\cite{PhysRevLett.107.127205}. The Hamiltonian in Eq.~\ref{eq: pseudo-spin Chern ins} describes a high-pseudo-spin Chern insulator in a thin film of square lattice, with the layer, spin, orbital and site indexed by $n, \alpha, \gamma, (i, j)$, respectively.

\begin{widetext}
\begin{align}
\label{eq: pseudo-spin Chern ins}
H^{0} & = t_k \sum_{\langle ij \rangle, n} c^{\dagger}_{i, n, \alpha, \gamma}  \, \mathbbm{1}_{s} \mathbbm{1}_{\tau} \, c_{j, n, \alpha', \gamma'} + i \lambda_{so} \sum_{\langle ij \rangle, n}  c^{\dagger}_{i, n, \alpha, \gamma} (\vec{s}_{\alpha \alpha'} \times \hat{d}_{ij}) \cdot \hat{z} \, \tau^z_{\gamma \gamma'} c_{j, n, \alpha', \gamma'}  \nonumber \\
    & \quad + t_{intra} \sum_{i, n}  c^{\dagger}_{i, n, \alpha, \gamma}  \, \mathbbm{1}_{s} \tau^x_{\gamma \gamma'} \, c_{i, n, \alpha', \gamma'}  
    + M \sum_{\langle \langle ij \rangle \rangle, n} c^{\dagger}_{i, n, \alpha, \gamma}  \, \mathbbm{1}_{s} \tau^x_{\gamma \gamma'} \, c_{j, n, \alpha', \gamma'} \nonumber \\
    & \quad + t_z \sum_{i, n} c^{\dagger}_{i, n, \alpha, \gamma}  \, \mathbbm{1}_{s} \tau^x_{\gamma \gamma'} \, c_{i, n+1, \alpha', \gamma'}.
\end{align}
\end{widetext}
The first term is nearest-neighbor in-plane hopping, the second term describes SOC with $\hat{d}_{ij}$ pointing from site $i$ to $j$, the third term is an intra-unit cell inter-orbital hopping, the fourth term is an in-plane next-nearest-neighbor hopping, and the last term couples the 2D layers together along the $z$-axis. The parameters are chosen such that the 3D system is a weak TI.

\begin{center}
\begin{figure}[h]
\includegraphics[scale = 0.28]{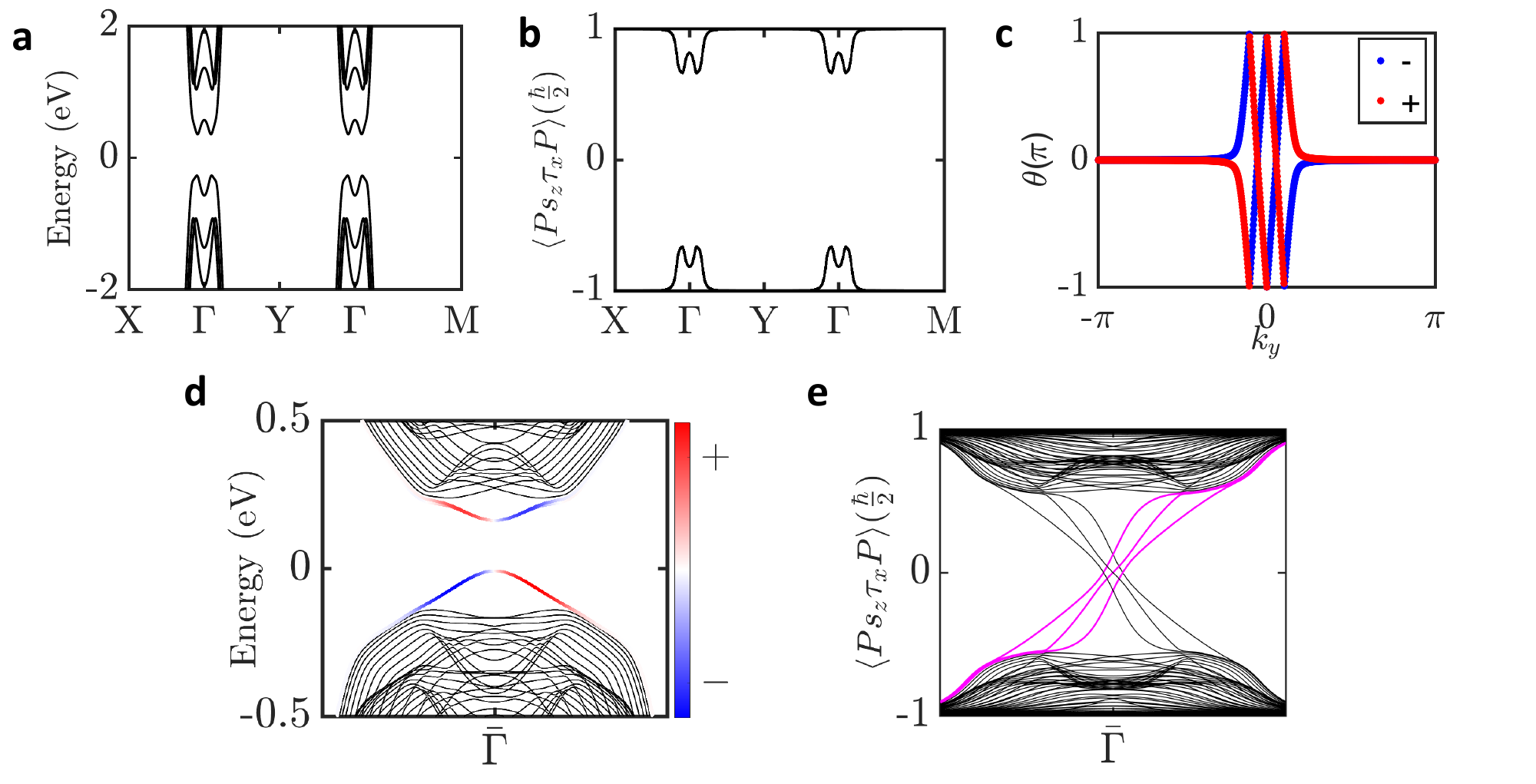}
\caption{\label{fig3} 
\textbf{Feature Chern Insulator: High-pseudo-spin Chern insulator.} In this work, we use \(t_k=-0.5\), \(\lambda_{SO}=-1\), \(M=7\), \(t_z=1\) and \(t_{intra}=-26.85\) for the model described in Eq.~\ref{eq: pseudo-spin Chern ins}.
(\textbf{a}) The band structure and (\textbf{b}) the feature spectrum \(\langle S_z\otimes\tau_x\otimes\sigma_0 \rangle\) of the high-pseudo-spin-Chern insulator, on which the feature spectrum topology can be defined since they are both gapped. 
(\textbf{c}) The Wilson loop for the two sectors in (\textbf{b}), where the +(-) indicates the sector with positive(negative) feature spectral values.
(\textbf{d}) The nanoribbon band structure of the high-pseudo-spin-Chern insulator along the \(\hat{y}\)-direction, in which the edge states on one edge are \( S_z\otimes\tau_x\otimes\sigma_0 \) resolved.
(\textbf{e}) The nanoribbon feature spectrum of the high-pseudo-spin-Chern insulator along the \(\hat{y}\)-direction with a Zeeman field in (\(\hat{x}+\hat{y}\))-direction. The edge feature spectra on the same edge in (\textbf{d}) are marked in magenta.}
\end{figure}
\end{center}

We consider a tri-layer thin film, whose band structures are shown in Fig.~\ref{fig3}(\textbf{a}). The chosen feature $s^z \otimes \tau^x \otimes \mathbbm{1}_{3 \times 3}$($s$, $\tau$, $\mathbbm{1}$ act on spin, orbital, layer space, respectively) captures the topology of the system, encoding more information than the $\mathds{Z}_2$ classification. The feature spectrum presented in Fig.~\ref{fig3}(\textbf{b}) consists of two separate branches, corresponding to the two feature sectors. The Wilson loop for the two sectors is plotted in Fig.~\ref{fig3}(\textbf{c}), from which we read the pseudo-spin Chern number is 3. As a result, the system is $\mathds{Z}_2$ non-trivial with a single Dirac cone on its edges(see Fig.~S1 in supplementary materials). Upon applying a weak Zeeman field, the Dirac cone becomes gapped, and meantime three gapless chiral edge states (due to three layers in the system) emerge in the edge feature spectrum( see Fig.~\ref{fig3}(\textbf{d, e}). This high pseudo-spin Chern insulators exhibits the feature-energy duality through its gapless edge energy or feature band structures. Moreover, it showcased that the feature spectrum topology is applicable even in the presence of a symmetry-breaking field.

Two more examples of feature Chern insulators can be found in the supplementary materials.

\section*{V. Feature Weyl Semimetals}
In the context of feature spectrum topology, besides the feature insulators discussed in last section, \emph{feature Weyl semimetals} constitute another class of topological materials\cite{PhysRevB.83.205101,https://doi.org/10.48550/arxiv.1301.0330}. As the name suggests, a feature Weyl semimetal hosts Weyl nodes in its bulk feature spectrum. The feature Weyl semimetal offers another interpretation of the topological insulators in 3D.

Using antiferromagnetic TI (AFM TI) \cite{PhysRevB.81.245209} as an example, where the non-trivial bulk topology is protected by a combination of time-reversal symmetry and a translation, we discuss an example of a feature Weyl semimetal, and how these feature Weyl nodes contain information about the topological phases in the system\cite{https://doi.org/10.48550/arxiv.2207.10099, Yang_2013}. 
\par
The AFM TI can be described by the Hamiltonian in Eq.~\ref{eq: pseudo-spin Chern ins}, $H_0$, with an additional staggered Zeeman field with field strength $h_z$,

\begin{equation}
\label{eq: AFM TI}
H^{\text{AFM TI}} = H^{0}\sigma^0 + h_z \sum_{i, n} c^{\dagger}_{i, n, \alpha, \gamma}  s^z_{\alpha \alpha'} \tau^z_{\gamma \gamma'}\sigma^z_{nn} \, c_{i, n, \alpha', \gamma'}.
\end{equation}

\begin{center}
\begin{figure}[h]
\includegraphics[scale=0.35]{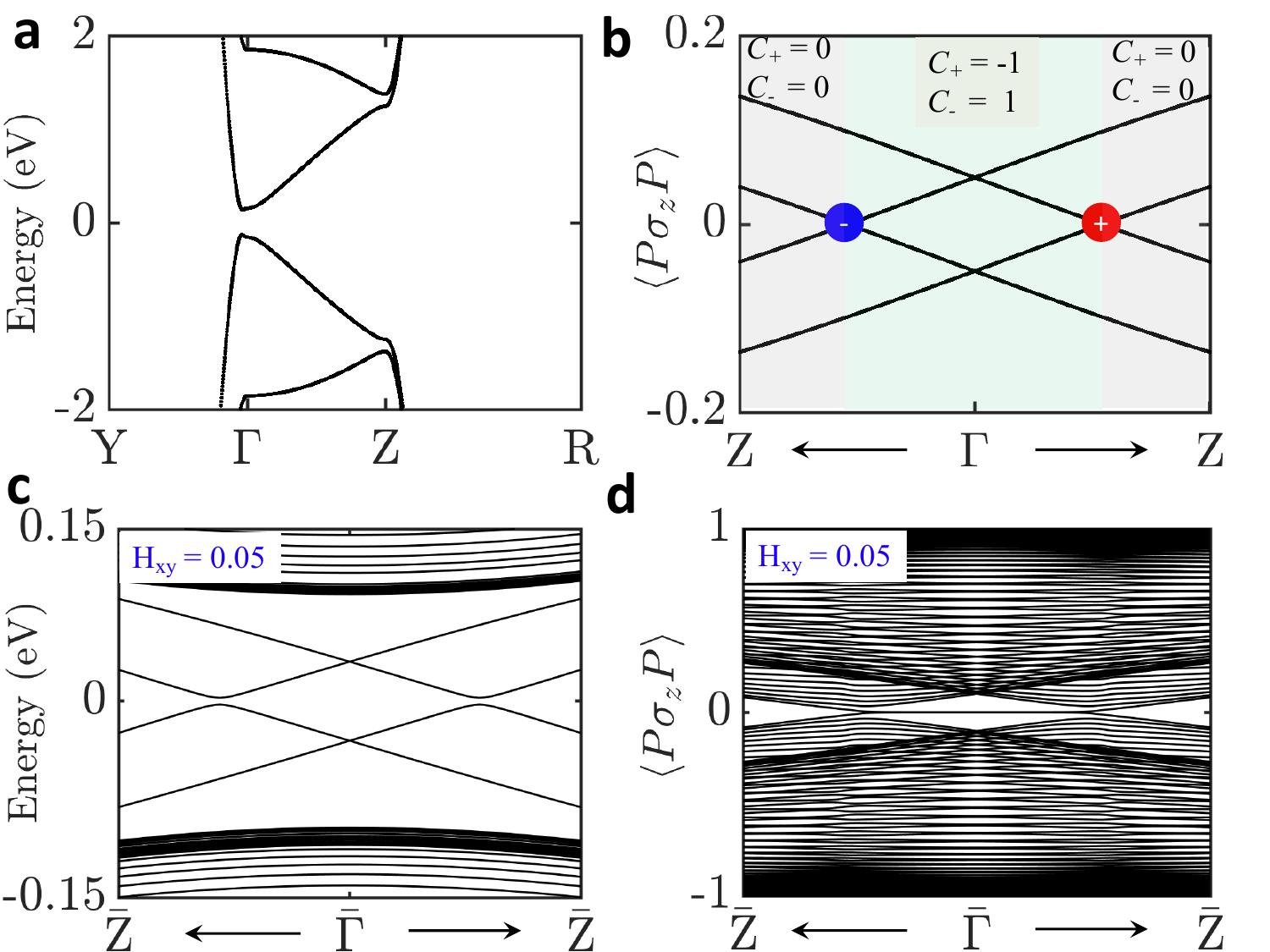}
\caption{\label{fig4} \textbf{Feature Weyl semimetals in the AFM TI.}
(\textbf{a}) The bulk energy band structure of the AFM TI, where $h_z=0.05$ is used for the model described in Eq.~\ref{eq: AFM TI}.
(\textbf{b}) The bulk feature spectrum of the AFM TI around \(\Gamma\) point. The red (blue) circle indicates the feature Weyl node with a positive (negative) feature chiral charge on the \(+\) sector. 
(\textbf{c}) The energy band structure of the $(010)$ slab of the AFM TI. An additional in-plane Zeeman field in (\(\hat{x}+\hat{y}\))-direction is added to gap the surface bands.
(\textbf{d}) The feature spectrum corresponds to (\textbf{c}). The surface Fermi arc in the feature spectrum connects the feature Weyl nodes. }
\end{figure}
\end{center}

With an appropriate choice of the parameters, the system is in a non-trivial AFM TI phase with fully gapped surface modes on the  $(001)$ surface and symmetry-protected gapless Dirac nodes on the $(100)$ and $(010)$ surfaces.,

The feature chosen here is $\mathbbm{1}_{4 \times 4} \otimes \sigma^z$, where $\sigma$ acts on the layer degree of freedom. Fig.~\ref{fig4}(\textbf{b}) shows that the two feature spectrum sectors (positive and negative), $\mathbbm{1}_{4 \times 4} \otimes \sigma^z = \pm 1$, have a non-trivial Chern number in the $k_z = 0$ slice, whilst the $k_z = \pi$ slice is trivial. This agrees with the $\mathds{Z}_2$ classification of AFM TI, which has a non-trivial $\mathds{Z}_2$ index in the $k_z = 0$ plane. Hence, there have to be two Weyl nodes with opposite chiral charges, located at $\pm k_{z,c}$ in the bulk feature spectrum, to account for the change of feature Chern number. This feature Weyl semimetal phase provides an alternative perspective of the AFM TI phases\cite{https://doi.org/10.48550/arxiv.2207.10099, Yang_2013}. (See Supplementary Discussion and Fig.~S6 and S7.)

Upon applying a symmetry-breaking Zeeman field, Fig.~\ref{fig4}(\textbf{c}) shows a gap opening up in the surface Dirac nodes on the (010) surface while the feature edge states are gapless Fig.~\ref{fig4}(\textbf{d}). The gapless feature edge states for the $\vert k_{z}\vert \leq k_{z,c}$ are analogous to Fermi arcs that connect Weyl nodes in 3D Weyl semimetals. 

\section*{VI. Role of Spin-Orbit Coupling}
All the previous discussions on feature insulators were focused on time-reversal odd features, which gave a feature Chern number classification of the ground-state topology. In this section, we propose a time-reversal invariant feature, $\mathbf{L\cdot S}$, which does not allow for a non-zero feature Chern number as the sectors of $P\mathbf{L\cdot S}P$ are time-reversal symmetric. We, therefore, have to revert to a $\mathds{Z}_2$-type classification based upon the feature Wilson loop to track the topological phase. As we show below, the feature spectrum is not only useful in characterizing non-conventional topological phases but also helpful for identifying the key component that induces the topological nature of a material.

\begin{center}
\begin{figure}[h]
\includegraphics[scale = 0.38]{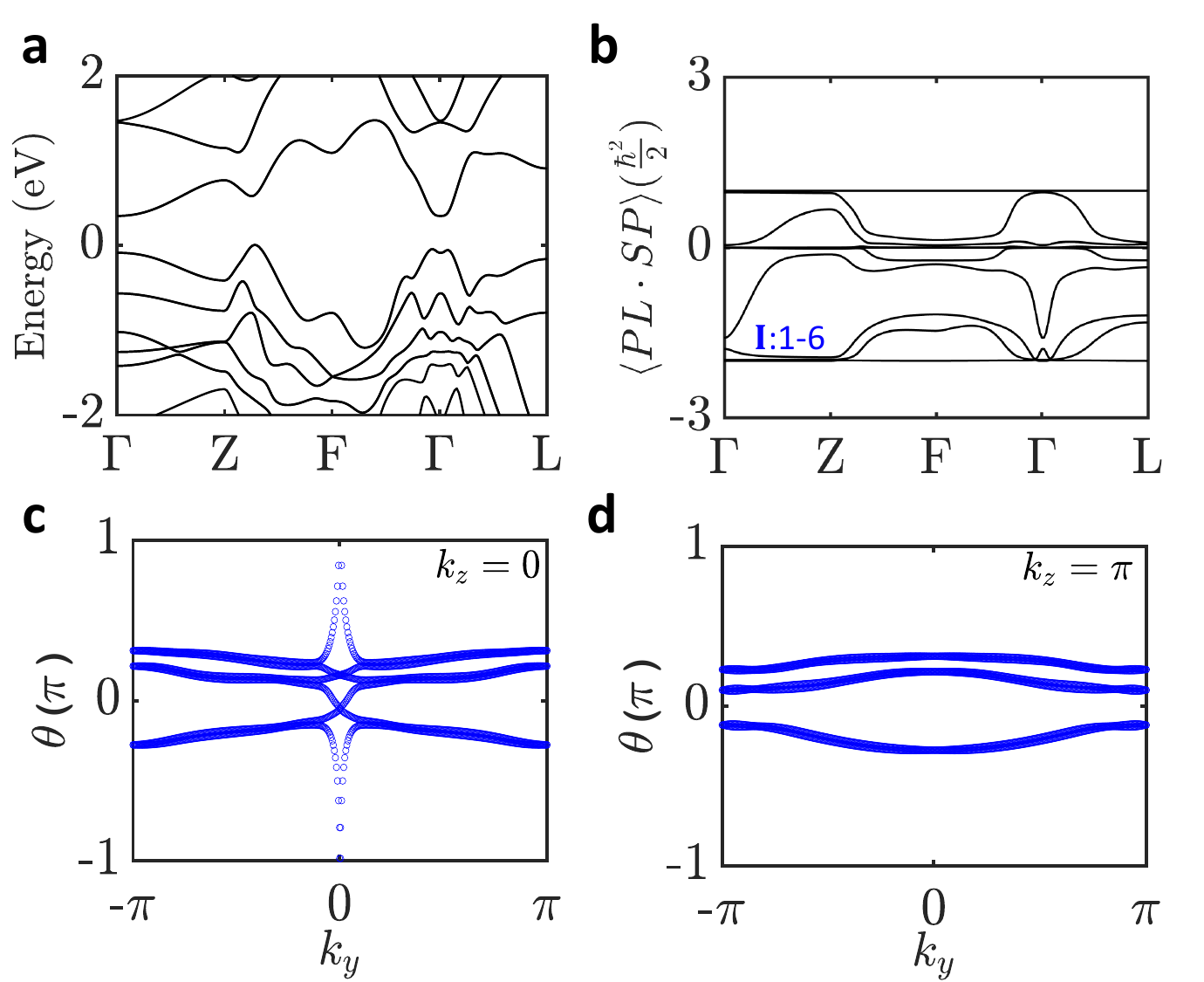}
\caption{\label{fig5} \textbf{Topological characterization of $\rm{Bi_2Se_3}$.}
(\textbf{a}) The bulk band structure of $\rm{Bi_2 Se_3}$ in the presence of SOC. (\textbf{b}) The bulk feature spectrum $\langle P\mathbf{L\cdot S}P\rangle$ of $\rm{Bi_2 Se_3}$ in the presence of SOC. The sector I includes the bands 1-6, as indicated in the figure. (\textbf{c, d}) The Wilson loop of the sector I on the $k_z=0$ and the $k_z=\pi$ plane, respectively.}
\end{figure}
\end{center}

We choose $\rm{Bi_2Se_3}$, a typical example of 3D TI, for this discussion. We employ a realistic tight-binding model that comprises the $s, p$ orbitals of  $\rm{Bi, Se}$ atoms\cite{PhysRevB.84.205424}. SOC causes a band-inversion at $\Gamma$ near the Fermi level (see Fig.~\ref{fig5}($\textbf{a}$)), and the SOC strength is varied to track the evolution of the band structures, with a band gap closing and reopening at 45\% SOC. Meanwhile, we track the evolution of the feature spectrum $P\mathbf{L\cdot S}P$. There are three sectors in the feature spectrum corresponding to the states $\vert J = \frac{1}{2}, L = 0, S = \frac{1}{2}\rangle$, $\vert J = \frac{1}{2}, L = 1, S = \frac{1}{2}\rangle$ and $\vert J = \frac{3}{2}, L = 1, S = \frac{1}{2}\rangle$, respectively. Our calculation finds that all three sectors are trivial before the band inversion for small SOC strength. After the band inversion, the sector associated with the feature $\vert J = \frac{1}{2}, L = 1, S = \frac{1}{2}\rangle$ becomes nontrivial. This sector, consisting of the feature bands 1-6, is labeled as sector I in Fig.~\ref{fig5}($\textbf{b}$). From the $k_y$-dependence of the feature Wannier center of this sector in the $k_z = 0$ and $k_z = \pi$ planes, we find that it is $\mathds{Z}_2$ topologically nontrivial, as shown in Fig.~\ref{fig5}($\textbf{c, d}$). It demonstrates that it is the bands in the sector $\vert J = \frac{1}{2}, L = 1, S = \frac{1}{2}\rangle$ that are responsible for the non-trivial topology of $\rm{Bi_2 Se_3}$ in the strong TI phase, in agreement with previous analysis \cite{zhang_liu_qi_dai_fang_zhang_2009}. This clearly shows that the spin-orbit feature spectrum $P\mathbf{L\cdot S}P$ is a more refined probe of the ground-state wave function topology.

\section*{VII. Discussion}
We have laid out the concept of feature spectrum topology based on projected quantum numbers, which we have demonstrated to be a more subtle and refined probe of the ground-state topology of quantum systems. The starting point of feature spectrum topology is to extend the topological consideration to the feature domain, aiming to establish a more comprehensive framework of topological characterization. The design principle lets the feature spectrum topology encompass the symmetry-based topological schemes (STS), like symmetry indicators and topological quantum chemistry. In other words, all the topological phases recognized by STS can be well incorporated into the feature spectrum; In reverse, the feature spectrum topology identifies much more topological phases that are invisible to STS, which can be understood in two aspects. 

On the one hand, the feature spectrum topology applies to the systems whose symmetry is broken, where STS fails to characterize the ground state topology. One issue identified in the study of TCIs is the lack of gapless states on boundaries that do not possess the required crystalline symmetry. We can apply the feature-energy duality here. When the energy band structure on the boundary is gapped, the feature spectrum for the boundary becomes gapless. This suggests that the topological boundary states cannot be completely removed but may be hidden within the bulk energy bands. This concept also applies to $\mathds{Z}_2$ TIs, where the boundary states may become gapped due to the introduction of magnetic ions, which breaks time-reversal symmetry. Despite this, the topological boundary states, responsible for the gapless feature spectrum on the boundary, must still exist in some form. Symmetry protection is not necessary for topological boundary states but rather indicates metallic behavior on the boundary bands. 
Moreover, the presence and the difference of mass terms in these gapped boundary states on different boundaries can give rise to soliton states at the lower dimensional boundary, giving the hinge or corner states via the mass-kink mechanism \cite{PhysRevLett.121.126402, PhysRevB.100.205126, PhysRevB.99.155102, arxiv2303.04031, khalaf_po_vishwanath_watanabe_2018, PhysRevB.97.205136}. Further introducing a symmetry-breaking term to the entire bulk can lead to the gap opening of all boundary energy bands and the invalidation of first-order topological invariants. In this scenario, higher-order topological invariants are introduced \cite{khalaf_po_vishwanath_watanabe_2018, PhysRevB.97.205136, doi:10.1126/science.aah6442, doi:10.1126/sciadv.aat0346, PhysRevB.97.205135, PhysRevLett.119.246402}. From the perspective of feature spectrum topology, it is natural to allow for such symmetry-breaking terms provided the bulk feature spectrum remains gapped, and the topology invariants associated with the feature sectors remain valid. The gapless feature spectrum on the boundary reflects the feature Wilson loop winding number $\mathds{Z}_O$ being a first-order topological invariant. Further research is needed to identify the appropriate feature operators for various higher-order and fragile topologies and to determine whether all higher-order TIs can be considered first-order in the feature spectrum topology. This work would provide a solid foundation for understanding the robustness of these systems from a topological perspective. 

On the other hand, the feature spectrum topology captures the topological phase whose topological band inversions are at generic $k$-points. In contrast, STS merely focuses on the band inversions at high-symmetry points, missing a great portion of materials with non-trivial ground state topology. The feature spectrum topology dramatically expands the pool of topological materials, providing more candidate materials with desirable properties. Previous research focused on the materials that host gapless edge bands to seek their applicability in electronics, spintronics, and other applications. The feature spectrum topology reminds us that even the phase without gapless boundary states can have such desirable properties due to their non-trivial bulk topology. Therefore, the importance of the feature-energy duality, as a manifestation of the bulk-boundary correspondence due to the non-trivial bulk topology, cannot be overemphasized.
Here, we illustrate this with two examples. Consider a SnTe thin film with an odd number of layers, which is shown to support two pairs of gapless Dirac cones on the edges protected by the mirror symmetry $M_z$. When built into a transistor, the gapless edge states can transport mirror-(spin-) polarized electrons. Applying a vertical electric field breaks the mirror symmetry and opens a gap in the edge bands, making the material seems less interesting \cite{Liu2014}. However, the feature-energy duality tells us that the thin film supports gapless edge feature bands that can still give the mirror-Hall effect with fine-tuned chemical potential, through which the electrons with different mirror polarizations accumulate on the two opposite boundaries (See Supplementary Discussion and Fig.~S4). Another example is the recently proposed $\alpha$-Sb, which was previously recognized as a topologically trivial insulator. Still, its non-trivial spin Chern number $\mathcal{C}_s$=2 leads to the emergence of edge bands, and hence the spin-accumulation through the nearly-quantized spin-Hall current with fine-tuned chemical potential, providing a platform for efficient spintronics\cite{sb} (also See Supplementary Materials and Fig.~S5). 
Similarly, a system can feature orbital Hall conductivity plateaus for the orbital Hall effect if the sectors in the feature spectrum $\mathbf{\hat{n}\cdot L}$ carry nonzero Chern numbers. Other cases can even have an isospin Hall conductivity plateaus if the sectors in the feature spectrum $\mathbf{\hat{n}\cdot}\boldsymbol{\tau}$ carry nonzero Chern numbers, where $\tau$ is the isospin indicating the bonding and anti-bonding states \cite{PhysRevB.82.165104, PhysRevB.81.115407}. In general, it applies to the feature Chern insulators featured with physical quantities $\hat{O}$, resulting in a generalized $\hat{O}$-Hall effect with measurable $\hat{O}$-conductivity plateaus \cite{PhysRevB.80.125327}. The feature spectrum topology provides opportunities for many materials that were previously recognized as topological trivial to support the feature-Hall effect, establishing a new type of electronic engineering, termed \emph{featuretronics}. 

Last but not least, the above example of $\rm{Bi_2Se_3}$ reveals another merit of the feature spectrum topology; that is, it provides a systematic and rigorous way to analyze which component plays a role in making the material topologically nontrivial, surpassing the intuitive but less rigorous methods such as band inversions. A thorough analysis of the feature spectrum thus helps to extract the essence of the topological phase in a material, paving the boulevard toward a more delicate and efficient topological material engineering.

\section*{Acknowledgments}
\textbf{Funding}: H.L. acknowledges the support of the National Science and Technology Council (NSTC) in Taiwan under grant number MOST 111-2112-M-001-057-MY3. The work at Northeastern University is supported by the Air Force Office of Scientific Research under award number FA9550-20-1-0322, and it benefited from computational resources of Northeastern University's Advanced Scientific Computation Center (ASCC) and the Discovery Cluster. 
\textbf{Author contributions:}H.L. designed research;  B.W., Y.-C.H., X.Z., T.O., and H.L. performed research; B.W., Y.-C.H., T.O., and H.L. analyzed data; and B.W., Y.-C.H., T.O., and H.L. wrote the paper.
\textbf{Competing interests:}The authors declare no competing interest.
\textbf{Data and materials availability:} All the data is available upon request.


\end{document}